\documentclass[aps,prd,superscriptaddress]{revtex4}
\usepackage{epsfig,epsf}
\usepackage{amsmath}
\usepackage{amsthm}
\usepackage{amsfonts}
\usepackage{amssymb}
\usepackage{dsfont}
\usepackage{multirow}
\usepackage{appendix}
\usepackage{slashed}
\usepackage[active]{srcltx}
\usepackage{psfrag}

\begin{document}

\title{The structure, mixing angle, mass and couplings of the light scalar $%
f_0(500)$ and $f_0(980)$ mesons}
\date{\today}
\author{S.~S.~Agaev}
\affiliation{Institute for Physical Problems, Baku State University, Az--1148 Baku,
Azerbaijan}
\author{K.~Azizi}
\affiliation{Department of Physics, Do\v{g}u\c{s} University, Acibadem-Kadik\"{o}y, 34722
Istanbul, Turkey}
\affiliation{School of Physics, Institute for Research in Fundamental Sciences (IPM),
P.~O.~Box 19395-5531, Tehran, Iran}
\author{H.~Sundu}
\affiliation{Department of Physics, Kocaeli University, 41380 Izmit, Turkey}

\begin{abstract}
The mixing angle, mass and couplings of the light scalar mesons $f_0(500)$
and $f_0(980)$ are calculated in the framework of QCD two-point sum rule
approach by assuming that they are tetraquarks with diquark-antidiquark
structures. The mesons are treated as mixtures of the heavy $|H\rangle
=([su][\bar s \bar u]+[sd][\bar s \bar d])/\sqrt 2$ and light $|L\rangle
=[ud][\bar u \bar d]$ scalar diquark-antidiquark components. We extract from
corresponding sum rules the mixing angles $\varphi_H$ and $\varphi_L$ of
these states and evaluate the masses and couplings of the particles $%
f_0(500) $ and $f_0(980)$.
\end{abstract}

\maketitle

\textbf{1. }\textit{Introduction.} Light scalar mesons that reside in the
region $m<1\mathrm{GeV}$ of the meson spectroscopy are sources of
long-standing problems for the conventional quark model. The standard
approach when treating mesons as bound states of a quark and an antiquark $q%
\bar{q}$ meets with evident troubles to include $f_{0}(500)$ and $f_{0}(980)$%
, as well as some other light particles into this scheme: There are
discrepancies between predictions of this model for a mass hierarchy of
light scalars and measured masses of these particles. Therefore, for
instance, the $f_{0}(980)$ meson was already considered as a four-quark
state with $q^{2}\bar{q}^{2}$ content \cite{Jaffe:1976ig}.

During passed decades  physicists made great efforts to understand features
of the light scalar mesons: They were   treated as meson-meson molecules
\cite{Weinstein:1982gc,Weinstein:1990gu,Achasov:1996ei,Branz:2007xp}, or
considered as diquark-antidiquark bound states \cite%
{Maiani:2004uc,Hooft:2008we}. These  models stimulated not only qualitative
analysis of the light scalar mesons, but also allowed one to calculate their
parameters using different methods. Thus, in Ref.\ \cite{Ebert:2008id}
masses of the $f_{0}(500)$, $f_{0}(980)$, $a_{0}(980)$ and $K_{0}^{\ast
}(800)$ mesons were evaluated in the context of the relativistic
diquark-antidiquark approach and nice agreements with the data were found.
There are growing understanding that the mesons from the  light scalars'
nonet are exotic particles or at least contain substantial multiquark
components: lattice simulations and experimental data seem support these
suggestions. Further information on relevant theoretical ideas and models, as well as
on experimental data can be found in original and review articles  \cite%
{Alford:2000mm,Amsler:2004ps,Bugg:2004xu,Jaffe:2004ph,Klempt:2007cp}.

Intensive studies of the light scalars as tetraquark states were carried out
using QCD sum rules method \cite%
{Latorre:1985uy,Narison:1986vw,Brito:2004tv,Wang:2005cn,
Chen:2007xr,Lee:2005hs,Sugiyama:2007sg,Kojo:2008hk,Wang:2015uha}. Essential
part of these investigations confirmed assignment of the light scalars as tetraquark
states despite the fact that to explain experimental data in some of them
authors had to introduce various modifications to a pure diquark-antidiquark
picture and to treat the particles as a mixture of diquark-antidiquarks with
different flavor structures \cite{Chen:2007xr}, or as superpositions of
diquark-antidiquark and $q\bar{q}$ components \cite%
{Sugiyama:2007sg,Kojo:2008hk,Wang:2015uha}. There was also the article (see,
Ref.\ \cite{Lee:2005hs}), results of which did not support an interpretation
of the light scalars as diquark-antidiquark bound states.

As is seen, theoretical analyses performed even within the same method
lead to different conclusions about the internal structures of the mesons
from the light scalar nonet. One should add to this picture also large
errors from which suffer experimental data on the masses and widths of these
particles \cite{Patrignani:2016xqp} to understand difficulty of existing
problems.

\textbf{2.} \textit{Mixing schemes. }An approach to the nonet of light
scalars as mixtures of tetraquarks belonging to different representations of
the color group was recently proposed in Ref.\ \cite{Kim:2017yvd}. In
accordance with this approach the nonet of the light spin-0 mesons can be
considered as tetraquarks composed of the color ($\overline{\mathbf{3}}_{c}$%
) and flavor ($\overline{\mathbf{3}}_{f}$) antitriplet scalar diquarks.
Then, these tetraquarks in the flavor space form a nonet of the scalar
particles $\overline{\mathbf{3}}_{f}\otimes \mathbf{3}_{f}=\mathbf{8}%
_{f}\oplus \mathbf{1}_{f}$. In order to embrace the second nonet of the
scalar mesons occupying the region above $1\ \mathrm{GeV}$ spin-1 diquarks
belonging to the color-flavor representation ($\mathbf{6}_{c},\ \overline{%
\mathbf{3}}_{f}$) can be used. The tetraquarks built of the spin-1 diquarks
have the same flavor structure as ones constructed from spin-0 diquarks, and
therefore can mix with them.

In the present Letter we restrict ourselves by considering only the first
nonet of the scalar particles. Therefore, in what follows we neglect their
possible mixing with tetraquarks composed of the spin-1 diquarks. The flavor
singlet and octet components of this nonet have the structures

\begin{eqnarray*}
|\mathbf{1}_{f}\rangle &=&\frac{1}{\sqrt{3}}\left\{ [su][\overline{s}%
\overline{u}]+[ds][\overline{d}\overline{s}]+[ud][\overline{u}\overline{d}%
]\right\} , \\
\,\,|\mathbf{8}_{f}\rangle &=&\frac{1}{\sqrt{6}}\left\{ [ds][\overline{d}%
\overline{s}]+[su][\overline{s}\overline{u}]-2[ud][\overline{u}\overline{d}%
]\right\} ,
\end{eqnarray*}%
that in the exact $SU_{f}(3)$ symmetry can be directly identified with the
physical mesons. But the real scalar particles are mixtures of these states,
and in the singlet-octet basis and one-angle mixing scheme have the
decomposition
\begin{equation}
\begin{pmatrix}
|f\rangle \\
|f^{\prime }\rangle%
\end{pmatrix}%
=U(\theta )%
\begin{pmatrix}
|\mathbf{1}_{f}\rangle \\
|\mathbf{8}_{f}\rangle%
\end{pmatrix}%
,\ \ U(\theta )=%
\begin{pmatrix}
\cos \theta & -\sin \theta \\
\sin \theta & \cos \theta%
\end{pmatrix}%
,
\end{equation}%
where, for the sake of simplicity, we denote $f=f_{0}(500)$ and $f^{\prime
}=f_{0}(980)$, and $\theta $ is the corresponding mixing angle.
Alternatively, one can introduce the heavy-light basis
\begin{equation}
|\mathbf{H}\rangle =\frac{1}{\sqrt{2}}\left\{ [su][\overline{s}\overline{u}%
]+[ds][\overline{d}\overline{s}]\right\} ,\ |\mathbf{L}\rangle =[ud][%
\overline{u}\overline{d}],
\end{equation}%
and for the physical mesons get the expansion
\begin{equation}
\begin{pmatrix}
|f\rangle \\
|f^{\prime }\rangle%
\end{pmatrix}%
=U(\varphi )%
\begin{pmatrix}
|\mathbf{H}\rangle \\
|\mathbf{L}\rangle%
\end{pmatrix}%
.
\end{equation}%
Here we use $\varphi $ as the mixing angle in the heavy-light basis. An
emerged situation is familiar to one from analysis of the mixing problems in
the nonet of the pseudoscalar mesons, namely in the $\eta -\eta ^{\prime }$
system \cite{Feldmann:1998vh,Feldmann:1998sh,Agaev:2014wna}. The heavy-light
basis in the case under consideration is similar to the quark-flavor basis
employed there. The mixing angles in the two basis are connected by the
simple relation%
\begin{equation}
\tan \theta =\frac{\sqrt{2}\tan \varphi +1}{\tan \varphi -\sqrt{2}}.
\label{eq:Angles}
\end{equation}
\ \qquad

In general, one may introduce also two-angles mixing scheme if it leads to a
better description of the experimental data
\begin{equation}
\begin{pmatrix}
|f\rangle \\
|f^{\prime }\rangle%
\end{pmatrix}%
=U(\varphi _{H,}\varphi _{L})%
\begin{pmatrix}
|\mathbf{H}\rangle \\
|\mathbf{L}\rangle%
\end{pmatrix}%
,U(\varphi _{H,}\varphi _{L})=%
\begin{pmatrix}
\cos \varphi _{H} & -\sin \varphi _{L} \\
\sin \varphi _{H} & \cos \varphi _{L}%
\end{pmatrix}%
.
\end{equation}%
The couplings in the $f-f^{\prime }$ system can be defined in the form
\begin{equation}
\langle 0|J^{i}|f(p)\rangle =F_{f}^{i}m_{f},\,\ \langle 0|J^{i}|f^{\prime
}(p)\rangle =F_{f^{\prime }}^{i}m_{f^{\prime }},\ \ i=H,L.  \label{eq:Coupl}
\end{equation}%
We suggest that the couplings follow pattern of state mixing in both one-
and two-angles scheme. In the general case of two-angles mixing scheme this
implies fulfillment of the equality
\begin{equation}
\begin{pmatrix}
F_{f}^{H} & F_{f}^{L} \\
F_{f^{\prime }}^{H} & F_{f^{\prime }}^{L}%
\end{pmatrix}%
=U(\varphi _{H,}\varphi _{L})%
\begin{pmatrix}
F_{H} & 0 \\
0 & F_{L}%
\end{pmatrix}%
,  \label{eq:2AngleCoupl}
\end{equation}%
where $F_{H}$ and $F_{L}$ may be formally interpreted as couplings of the
"particles" $|\mathbf{H}\rangle $ and $|\mathbf{L}\rangle .$

Currents $J^{H}(x)$ and $J^{L}(x)$ in Eq.\ (\ref{eq:Coupl}) that correspond
to $|\mathbf{H}\rangle $ and $|\mathbf{L}\rangle $ states are given by the
expressions%
\begin{equation}
J^{H}(x)=\frac{\epsilon ^{dab}\epsilon ^{dce}}{\sqrt{2}}\left\{ \left[
u_{a}^{T}(x)C\gamma _{5}s_{b}(x)\right] \left[ \overline{u}_{c}(x)\gamma
_{5}C\overline{s}_{e}^{T}(x)\right] +\left[ d_{a}^{T}(x)C\gamma _{5}s_{b}(x)%
\right] \left[ \overline{d}_{c}(x)\gamma _{5}C\overline{s}_{e}^{T}(x)\right]
\right\} ,  \label{eq:Curr1}
\end{equation}%
and
\begin{equation}
J^{L}(x)=\epsilon ^{dab}\epsilon ^{dce}\left[ u_{a}^{T}(x)C\gamma
_{5}d_{b}(x)\right] \left[ \overline{u}_{c}(x)\gamma _{5}C\overline{d}%
_{e}^{T}(x)\right] ,  \label{eq:Curr2}
\end{equation}%
where $a,b,c,d$ and $e$ are color indices and $C$ is the charge conjugation
operator. Then the interpolating currents for physical states $J^{f}(x)$ and
$J^{f^{\prime }}(x)$ take the forms
\begin{equation}
\begin{pmatrix}
J^{f}(x) \\
J^{f^{\prime }}(x)%
\end{pmatrix}%
=U(\varphi _{H,}\varphi _{L})%
\begin{pmatrix}
J^{H}(x) \\
J^{L}(x)%
\end{pmatrix}%
.  \label{eq:Curr3}
\end{equation}%
In the simple case of one-angle mixing scheme Eq.\ (\ref{eq:Curr3})
transforms to the familiar superpositions%
\begin{equation}
J^{f}(x)=J^{H}(x)\cos \varphi -J^{L}(x)\sin \varphi ,\ \ \ \ J^{f^{\prime
}}(x)=J^{H}(x)\sin \varphi +J^{L}(x)\cos \varphi .  \label{eq:Curr4}
\end{equation}
These currents or their more complicated forms in the two-angles mixing
scheme may be used in QCD sum rule calculations to evaluate the masses and
couplings of the mesons $f$ and $f^{\prime }$.

\textbf{3. } \textit{Sum rules.} At the first stage of our calculations we
derive the sum rule for the mixing angle $\varphi $ of the $f-f^{\prime }$
system. To this end, we use the heavy-light basis and one-angle mixing
scheme and start from the correlation function \cite{Aliev:2010ra}
\begin{equation}
\Pi (p)=i\int d^{4}xe^{ip\cdot x}\langle 0|\mathcal{T}\{J^{f}(x)J^{f^{\prime
}\dagger }(0)\}|0\rangle .  \label{eq:CorrF1}
\end{equation}%
The sum rule obtained using $\Pi (p)$ allow us to fix the mixing angle $%
\varphi $. In fact, because the currents $J^{f}(x)$ and $J^{f^{\prime }}(x)$
create only $|f\rangle $ and $|f^{\prime }\rangle $ mesons, respectively, a
phenomenological expression for the correlator $\Pi ^{\mathrm{Phys}}(p)$
equals to zero. Then the second ingredient of the sum rule, namely
expression of the correlation function calculated in terms of quark-gluon
degrees of freedom $\Pi ^{\mathrm{OPE}}(p)$ should be equal to zero. Because
$\Pi ^{\mathrm{OPE}}(p)$ depends on the mixing angle $\varphi ,$ it is not
difficult to find%
\begin{equation}
\tan 2\varphi =\frac{2\Pi _{HL}^{\mathrm{OPE}}(p)}{\Pi _{LL}^{\mathrm{OPE}%
}(p)-\Pi _{HH}^{\mathrm{OPE}}(p)},  \label{eq:Angle}
\end{equation}%
where
\begin{equation}
\Pi _{ij}(p)=i\int d^{4}xe^{ip\cdot x}\langle 0|\mathcal{T}%
\{J^{i}(x)J^{j\dagger }(0)\}|0\rangle .
\end{equation}%
In deriving of Eq.\ (\ref{eq:Angle}) we benefited from the fact that $\Pi
_{HL}^{\mathrm{OPE}}(p)=\Pi _{LH}^{\mathrm{OPE}}(p),$ which can be proved by
explicit calculations. After applying the Borel transformation and
performing required continuum subtractions one can employ it to evaluate $%
\varphi $.

Having found the mixing angle we proceed and evaluate the spectroscopic
parameters of the mesons $f$ and $f^{\prime }$. The correlation functions
\begin{equation}
\Pi _{f}(p)=i\int d^{4}xe^{ip\cdot x}\langle 0|\mathcal{T}%
\{J^{f}(x)J^{f\dagger }(0)\}|0\rangle ,\ \ \Pi _{f^{\prime }}(p)=i\int
d^{4}xe^{ip\cdot x}\langle 0|\mathcal{T}\{J^{f^{\prime }}(x)J^{f^{\prime
}\dagger }(0)\}|0\rangle ,  \label{eq:CorrF2}
\end{equation}%
are appropriate for these purposes and can be utilized to derive the
relevant sum rules. The expression of $\Pi _{f}(p)$ in terms of the physical
parameters of the $f$ meson is given by the following simple formula%
\begin{equation*}
\Pi _{f}^{\mathrm{Phys}}(p)=\frac{\langle 0|J^{f}|f(p)\rangle \langle
f(p)|J^{f\dagger }|0\rangle }{m_{f}^{2}-p^{2}}+\ldots ,
\end{equation*}%
where the dots stand for contributions of the higher resonances and
continuum states. Using the interpolating current and matrix elements of the
$f$ meson from Eqs.\ (\ref{eq:Curr4})\ and (\ref{eq:Coupl}) it is a easy
task to show that
\begin{equation*}
\Pi _{f}^{\mathrm{Phys}}(p)=\frac{m_{f}^{2}}{m_{f}^{2}-p^{2}}\left(
F_{H}\cos ^{2}\varphi +F_{L}\sin ^{2}\varphi \right) ^{2}+\ldots .
\end{equation*}%
After calculating the correlation function $\Pi _{f}^{\mathrm{OPE}}(p)$ and
applying the Borel transformation in conjunction with continuum subtraction
one gets the sum rule
\begin{equation}
m_{f}^{2}\left( F_{H}\cos ^{2}\varphi +F_{L}\sin ^{2}\varphi \right)
^{2}e^{-m_{f}^{2}/M^{2}}=\Pi _{f}(s_{0},M^{2}),  \label{eq:SR1}
\end{equation}%
where $\Pi _{f}(s_{0},M^{2})=\mathcal{B}$ $\Pi _{f}^{\mathrm{OPE}}(p)$ is
the Borel transformed and subtracted expression of $\Pi _{f}^{\mathrm{OPE}%
}(p)$ with $M^{2}$ and $s_{0}$ being the Borel and continuum threshold
parameters, respectively. This sum rule and another one obtained from Eq.\ (%
\ref{eq:SR1}) by means of the standard operation $d/d(-1/M^{2})$ can be used
to evaluate the mass of the $f$ meson.

The similar analysis for $f^{\prime }$ yields
\begin{equation}
m_{f^{\prime }}^{2}\left( F_{H}\sin ^{2}\varphi +F_{L}\cos ^{2}\varphi
\right) ^{2}e^{-m_{f^{\prime }}^{2}/M^{2}}=\Pi _{f^{\prime }}(s_{0},M^{2}).
\label{eq:SR2}
\end{equation}%
From Eqs.\ (\ref{eq:SR1}) and (\ref{eq:SR2}) it is also possible to extract $%
\left( F_{H}\cos ^{2}\varphi +F_{L}\sin ^{2}\varphi \right) ^{2}$ and $%
\left( F_{H}\sin ^{2}\varphi +F_{L}\cos ^{2}\varphi \right) ^{2}$ for
evaluating of the couplings $F_{H}$ and $F_{L}$, but they may suffer from
large uncertainties: We instead evaluate $F_{H}$ and $F_{L}$ from sum rules
for the scalar "particles" $|\mathbf{H}\rangle $ and $|\mathbf{L}\rangle $,
using to this end correlation functions $\Pi _{H}(p)$ and $\Pi _{L}(p)$
given by Eq. (\ref{eq:CorrF2}), where $J^{f}(x)$ and $J^{f^{\prime }}(x)$
are replaced by $J^{H}(x)$ and $J^{L}(x)$, respectively.

\textbf{4. \ }\textit{Numerical results. } In calculations we utilize the
light quark propagator
\begin{eqnarray}
&&S_{q}^{ab}(x)=i\delta _{ab}\frac{\slashed x}{2\pi ^{2}x^{4}}-\delta _{ab}%
\frac{m_{q}}{4\pi ^{2}x^{2}}-\delta _{ab}\frac{\langle \overline{q}q\rangle
}{12}+i\delta _{ab}\frac{\slashed xm_{q}\langle \overline{q}q\rangle }{48}%
-\delta _{ab}\frac{x^{2}}{192}\langle \overline{q}g_{s}\sigma Gq\rangle
+i\delta _{ab}\frac{x^{2}\slashed xm_{q}}{1152}\langle \overline{q}%
g_{s}\sigma Gq\rangle  \notag \\
&&-i\frac{g_{s}G_{ab}^{\alpha \beta }}{32\pi ^{2}x^{2}}\left[ \slashed x{%
\sigma _{\alpha \beta }+\sigma _{\alpha \beta }}\slashed x\right] -i\delta
_{ab}\frac{x^{2}\slashed xg_{s}^{2}\langle \overline{q}q\rangle ^{2}}{7776}%
-\delta _{ab}\frac{x^{4}\langle \overline{q}q\rangle \langle
g_{s}^{2}G^{2}\rangle }{27648}+\ldots ,  \label{eq:qprop}
\end{eqnarray}%
and take into account quark, gluon and mixed operators up to dimension
twelve. The vacuum expectations values of the operators used in numerical
computations are well known: $\langle \bar{q}q\rangle =-(0.24\pm 0.01)^{3}\
\mathrm{GeV}^{3}$, $\langle \bar{s}s\rangle =0.8\ \langle \bar{q}q\rangle $,
$\langle \overline{q}g_{s}\sigma Gq\rangle =m_{0}^{2}\langle \overline{q}%
q\rangle $, $\langle \overline{s}g_{s}\sigma Gs\rangle =m_{0}^{2}\langle
\bar{s}s\rangle $, $\langle \alpha _{s}G^{2}/\pi \rangle =(0.012\pm 0.004)\,%
\mathrm{GeV}^{4}$, $\langle g_{s}^{3}G^{3}\rangle =(0.57\pm 0.29)\ \mathrm{%
GeV}^{6}$, where $m_{0}^{2}=(0.8\pm 0.1)\ \mathrm{GeV}^{2}$.

The working regions for the Borel and continuum threshold parameters are
fixed in the following form%
\begin{equation}
M^{2}=(1.1-1.3)\ \mathrm{GeV}^{2},\ \ s_{0}=(1.4-1.6)\ \mathrm{GeV}^{2},
\label{eq:Regions}
\end{equation}%
that satisfy standard requirements of sum rules computations. For example,
at the lower limit of the Borel parameter the sum of the dimension-10, 11
and 12 terms in $\Pi _{LL}(s_{0},M^{2})-\Pi _{HH}(s_{0},M^{2})$ does not
exceed $5\%$ of all contributions. At the upper bound of the working window
for $M^{2}$ the pole contribution to the same quantity is larger than $0.12$
of the whole result, which is typical for multiquark systems. Variation of
the auxiliary parameters $M^{2}$ and $s_{0}$ within the regions (\ref%
{eq:Regions}), as well as uncertainties of the other input parameters
generate theoretical errors of sum rules computations. The $\tan 2\varphi $
extracted using Eq.\ (\ref{eq:Angle}), as is seen from Fig.1 (a),
demonstrates a mild dependence on $M^{2}.$ As a result, it is not difficult
to estimate that
\begin{equation}
\varphi =-27^{\circ }.66\pm 1^{\circ }.24.  \label{eq:AngRes}
\end{equation}%
This value of $\varphi $ in the heavy-light basis is equivalent to $\theta
=-33^{\circ }.00\pm 1^{\circ }.17$ in the singlet-octet basis. Using Eq.\ (%
\ref{eq:AngRes}) it is not difficult to evaluate the mesons' masses in the
one-angle mixing scheme that read
\begin{equation}
m_{f}=(597\pm 81)\ \ \mathrm{MeV,\ \ \ \ }m_{f^{\prime }}=(902\pm 125)\ \ \
\mathrm{MeV.}  \label{eq:Masses1A}
\end{equation}%
As is seen, the one-angle mixing scheme, if take into account the central
values from Eq.\ (\ref{eq:Masses1A}), does not describe correctly the
experimental data: it overshoots the mass of the $f_{0}(500)$ meson and, at
the same time, underestimates the mass of the $f_{0}(980)$ meson. The
agreement can be improved by introducing the two mixing angles $\varphi _{H}$
and $\varphi _{L}$. It turns out that to achieve a nice agreement with the
available experimental data it is enough to vary $\varphi _{H}$ and $\varphi
_{L}$ within the limits (\ref{eq:AngRes}):
\begin{equation}
\varphi _{H}=-28^{\circ }.87\pm 0^{\circ }.42,\ \ \varphi _{L}=-\ 27^{\circ
}.66\pm 0^{\circ }.31.\ \   \label{eq:AngRes1}
\end{equation}%
For $m_{f}$ and $m_{f^{\prime }}$ the sum rules with two mixing angles $%
\varphi _{H}$ and $\ \varphi _{L}$ lead to predictions
\begin{equation}
m_{f}=(518\pm 74)\ \ \mathrm{MeV,\ \ \ \ }m_{f^{\prime }}=(996\pm 130)\ \ \
\mathrm{MeV,}  \label{eq:MassRes}
\end{equation}%
which are compatible with experimental data. The theoretical errors in Eq.\ (%
\ref{eq:MassRes}) accumulate uncertainties connected with $M^{2}$ and $s_{0}$%
, as well as arising from other input parameters. The dependence of $m_{f}$
and $m_{f^{\prime }}$ on the auxiliary parameters $M^{2}$ and $s_{0}$ does
not exceed limits allowed for such kind of calculations: In Figs. 1(b) and
1(c) we plot $m_{f}$ and $m_{f^{\prime }}$ as functions of the Borel
parameter to confirm a stability of corresponding sum rules.

In the two-angle mixing scheme the system of the physical particles $%
f-f^{\prime }$ is characterized by four couplings (\ref{eq:2AngleCoupl}).
After determining the mixing angles $\varphi _{H}$ and $\varphi _{L}$ that
fix the matrix $U(\varphi _{H,}\varphi _{L})$, quantities which should be
found from the relevant sum rules are only the couplings $F_{H}$ and $F_{L}$%
. As we have mentioned above to this end we consider two additional sum
rules by treating basic states $|\mathbf{H}\rangle $ and $|\mathbf{L}\rangle
$ as real "particles" and obtain%
\begin{equation}
F_{H}=(1.35\pm 0.34)\cdot 10^{-3}\ \mathrm{GeV}^{4},\ \ \ F_{L}=(0.68\pm
0.17)\cdot 10^{-3}\ \mathrm{GeV}^{4}.  \label{eq:CouplRes}
\end{equation}%
The coupling $F_{H}$ calculated in the present work is comparable with one
found in Ref. \ \cite{Brito:2004tv} using the same interpolating current (%
\ref{eq:Curr1}) and vacuum condensates up to dimension six and is given by $%
F_{H}=(1.51\pm 0.14)\cdot 10^{-3}\ \mathrm{GeV}^{4}$.

The mixing angles $(\varphi _{H},\ \varphi _{L})$, the masses $(m_{f},\
m_{f^{\prime }})$ and the couplings $(F_{H},\ F_{L})$ complete the set of
the spectroscopic parameters of the $f_0(500)$ and $f_0(980)$ mesons.
\begin{figure}[h!]
\begin{center}
\includegraphics[totalheight=5cm,width=5.5cm]{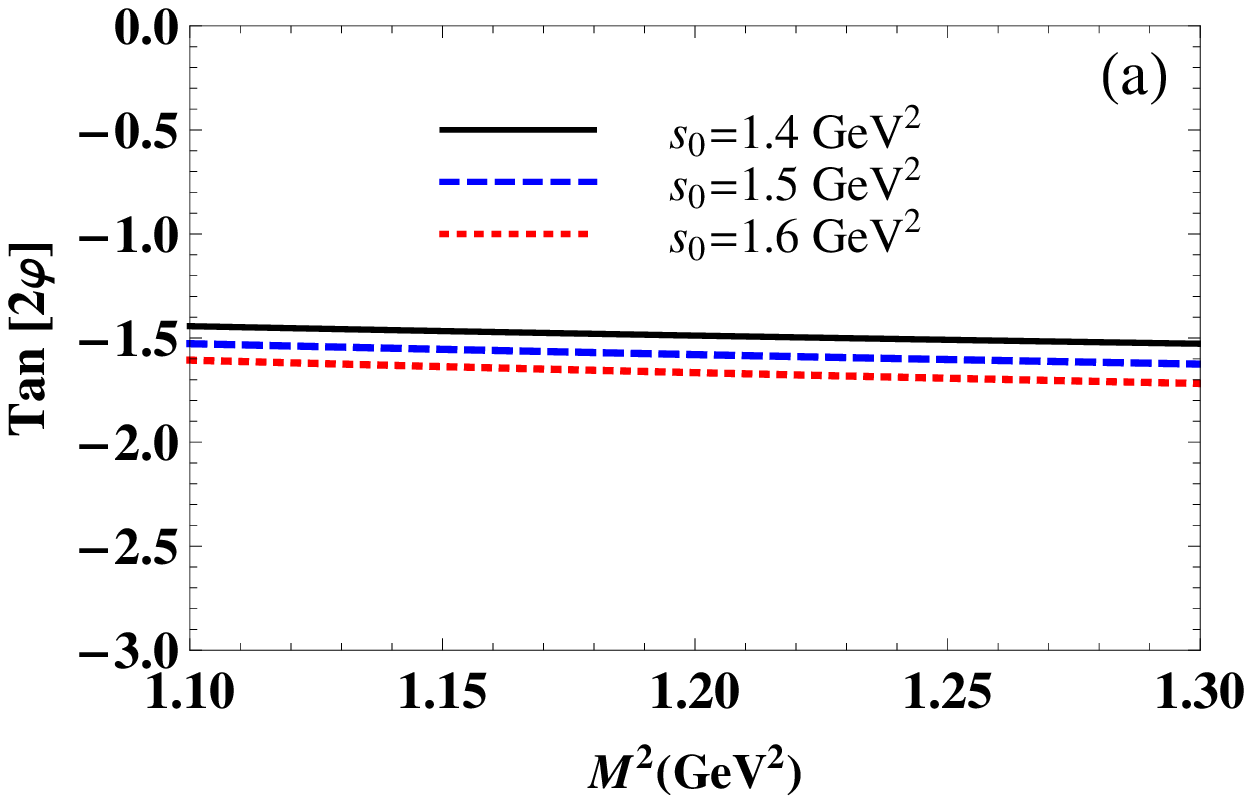}\,\, %
\includegraphics[totalheight=5cm,width=5.5cm]{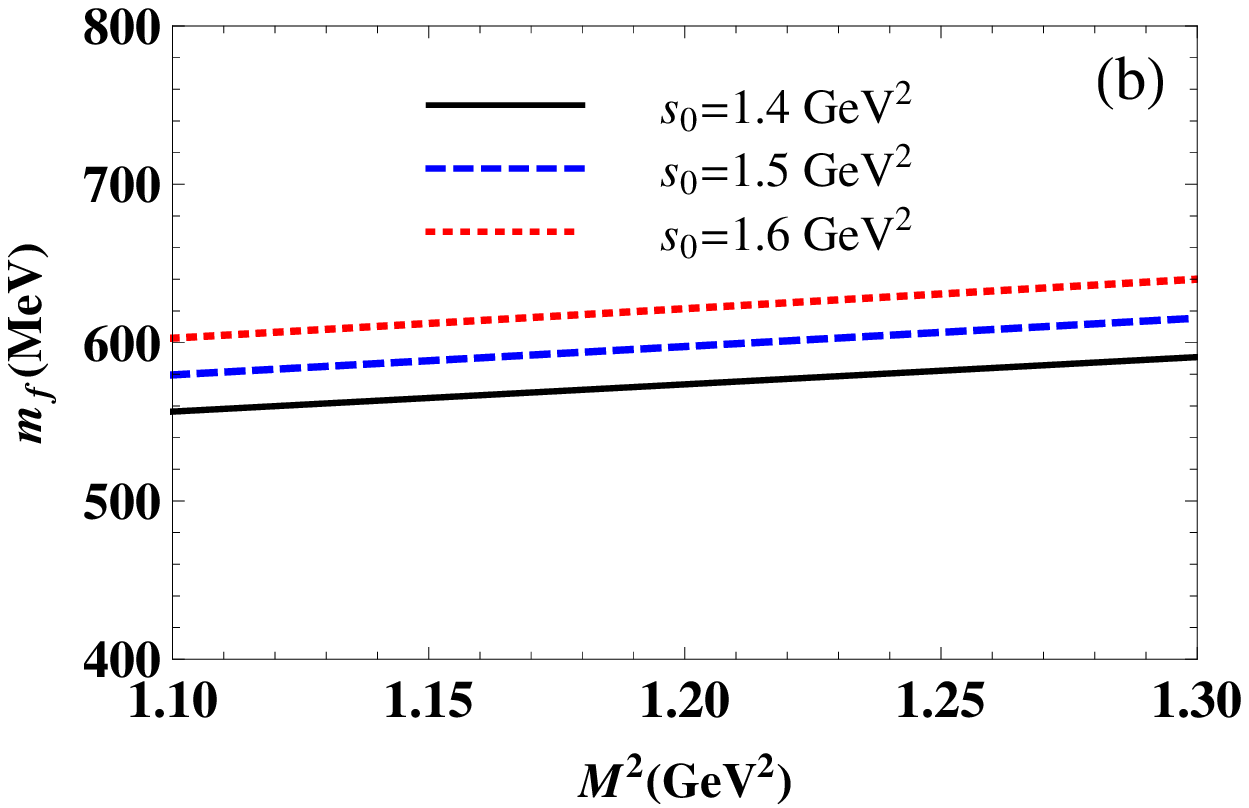} %
\includegraphics[totalheight=5cm,width=5.5cm]{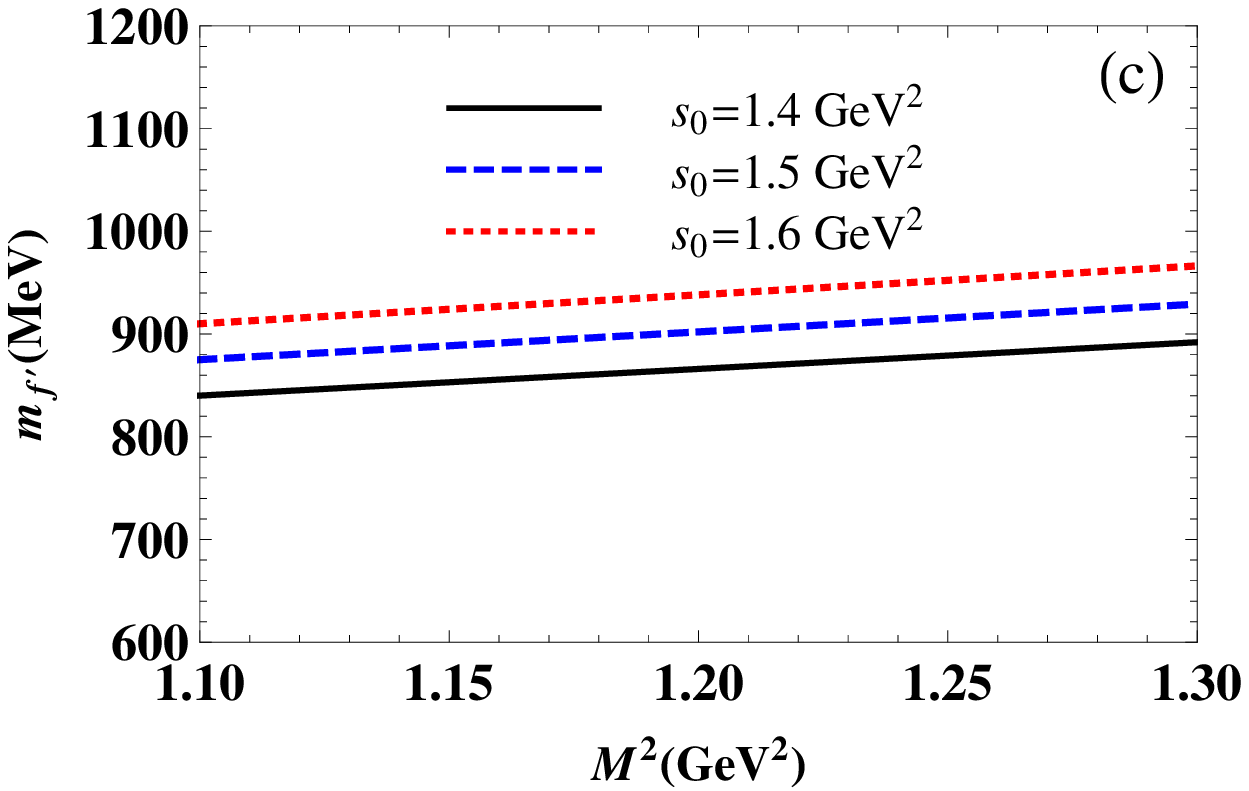}
\end{center}
\caption{ The $\tan 2\protect\varphi$ (a), and the masses $m_{f}$
(b) and $m_{f^{\prime}}$ (c) in the two-angles mixing scheme as
functions of the Borel parameter $M^2$ at fixed $s_0$ .}
\label{fig:SRresults}
\end{figure}

\textbf{5. } \textit{Concluding notes. } The investigation performed in the
present Letter has allowed us to calculate the mass and couplings of the $%
f_{0}(500)$ and $f_{0}(980)$ mesons by treating them as the mixtures of the
diquark-antidiquarks $|\mathbf{H}\rangle $ and $|\mathbf{L}\rangle $. We
have demonstrated that by choosing the heavy-light basis and mixing angles $%
\varphi _{H}=-28^{\circ }.87\pm 0^{\circ }.42$ and $\ \varphi _{L}=-\
27^{\circ }.66\pm 0^{\circ }.31$ a reasonable agreement with experimental
data can be achieved even information on the $f_{0}(500)$ meson suffers from
large uncertainties \cite{Patrignani:2016xqp}. The assumption on structures
of the light mesons made in the present work determines also their possible
decay mechanisms. Indeed, it is known that the dominant decay channels of
the $f_{0}(500)$ and $f_{0}(980)$ mesons are $f_{0}(500)\rightarrow \pi \pi $
and $f_{0}(980)\rightarrow \pi \pi $ processes. In experiments the decay $%
f_{0}(980)\rightarrow K\overline{K}$ was seen, as well. The mixing of the $|%
\mathbf{H}\rangle $ and $|\mathbf{L}\rangle $ diquark-antidiquark states to
form the physical mesons implies that all of these decays can run through
the superallowed Okubo-Zweig-Iizuka (OZI) mechanism: Without the mixing the
decay $f_{0}(980)\rightarrow \pi \pi $ can proceed due to one gluon
exchange, whereas $f_{0}(980)\rightarrow K\overline{K}$ is still OZI
superallowed process \cite{Brito:2004tv}. The another problem that finds its
natural explanation within the mixing framework is a large difference
between the full width of the mesons $f_{0}(500)$ and $%
f_{0}(980)$, which amount to $\Gamma=400-700\ \mathrm{MeV}$ and
$\Gamma=10-100\ \mathrm{MeV}$ \cite{Patrignani:2016xqp}, respectively.
In fact, the strong couplings $g_{f_{0}(500)\pi \pi }$ and $g_{f_{0}(980)\pi \pi }$
that determine the
width of the dominant partial decays $f_{0}(500)\rightarrow \pi \pi $ and $%
f_{0}(980)\rightarrow \pi \pi $ depend on the mixing angle $\varphi _{L}$
in the form
\begin{equation}
g^2_{f_{0}(500)\pi \pi }\propto \frac{1}{\sin ^{2}\varphi _{L}},\ \ \
g^2_{f_{0}(980)\pi \pi }\propto \frac{1}{\cos ^{2}\varphi _{L}}.
\end{equation}%
In the model under consideration this dependence is a main source that generates the numerical
difference between the partial width of aforementioned processes, and hence between
the full width of the mesons $f_{0}(500)$ and $f_{0}(980)$.

Analysis of the partial decays of the mesons $f_{0}(500)$ and $f_{0}(980)$, as well as
calculation of the spectroscopic parameters of other light scalar mesons
deserves further detailed investigations results of which will be published
elsewhere.\\

K.~A.~ thanks T\"{U}%
BITAK for the partial financial support provided under
Grant No. 115F183.

\end{document}